# Does the public discuss other topics on climate change than researchers?

# A comparison of explorative networks based on author keywords and hashtags

Robin Haunschild*[+], Loet Leydesdorff**, Lutz Bornmann***, Iina Hellsten**, Werner Marx*

[+] Corresponding author

* Max Planck Institute for Solid State Research

Heisenbergstraße 1, 70569 Stuttgart, Germany

Email: r.haunschild@fkf.mpg.de, w.marx@fkf.mpg.de

** Amsterdam School of Communication Science (ASCoR)

University of Amsterdam, PB 15793, 1001 NG Amsterdam, The Netherlands

E-mail. loet@leydesdorff.net, i.r.hellsten@uva.nl

*** Administrative Headquarters of the Max Planck Society

Division for Science and Innovation Studies

Hofgartenstr. 8, 80539 Munich, Germany.

Email: bornmann@gv.mpg.de

**Abstract**

Twitter accounts have already been used in many scientometric studies, but the meaningfulness of the data for societal impact measurements in research evaluation has been questioned. Earlier research focused on social media counts and neglected the interactive nature of the data. We explore a new network approach based on Twitter data in which we compare author keywords to hashtags as indicators of topics. We analyze the topics of tweeted publications and compare them with the topics of all publications (tweeted and not tweeted). Our exploratory study is based on a comprehensive publication set of climate change research. We are interested in whether Twitter data are able to reveal topics of public discussions which can be separated from research-focused topics. We find that the most tweeted topics regarding climate change research focus on the consequences of climate change for humans. Twitter users are interested in climate change publications which forecast effects of a changing climate on the environment and to adaptation, mitigation and management issues rather than in the methodology of climate-change research and causes of climate change. Our results indicate that publications using scientific jargon are less likely to be tweeted than publications using more general keywords. Twitter networks seem to be able to visualize public discussions about specific topics.

# 1 Introduction

In the context of an increasing accountability of the science sector in the public sphere, the demonstration of societal impact (popularity, attention, visibility, etc.) of research has become a new work area of scientometrics (Bornmann, 2016; Bornmann & Haunschild, 2017). Following the definition of Mohammadi, Thelwall, and Kousha (2016, p. 1198), we consider societal impact, in a broad sense, as “all types of research impact inside and outside of academia.” One is interested in finding metrics which can be used for measuring the impact of research on other sectors of society than science (in other words, one is interested in a complementary metric to citations which measures the impact of papers on the scientific sector).

Since the advent of altmetrics in the scientometrics universe, a new set of metrics, such as mentions and downloads of papers, has been proposed as potential indicators for research evaluation, and has been advocated as a potential indicator of societal impact (Sugimoto, Work, Larivière, & Haustein, 2017, p. 2046). According to the National Information Standards Organization (2016), “altmetrics may offer insight into impact by calculating an output’s reach, social relevance, and attention from a given community, which may include members of the public sphere” (p. 1). Altmetric indicators have been introduced in a manifesto (Priem, Taraborelli, Groth, & Neylon, 2010) and are defined as follows: “altmetric indicators estimate research impact by quantifying the dissemination of scholarly output in social media. Examples include mentions in blogs, number of tweets and retweets or inclusion in social bookmarking services” (Pooladian & Borrego, 2016, p. 1136).

Moed (2017) distinguishes four types of altmetrics: social media (e.g., Twitter and Facebook), reference managers (e.g., Mendeley), various forms of scholarly blogs, and coverage in mass media (e.g., daily newspapers). Mas-Bleda and Thelwall (2016) compared altmetrics for academic papers, and found that almost 80% of the papers received at least one read in Mendeley and more than a third (34%) of the academic papers were mentioned on Twitter. Overviews of studies dealing with altmetrics can be found in Bornmann (2014) and Sugimoto, Work, Larivière, and Haustein (2016). González-Valiente, Pacheco-Mendoza, and Arencibia-Jorge (2016) found more than 250 documents published between 2005 and 2015 in this area. Meanwhile, many publishers add social media metrics to their publications, such as Elsevier, Wiley, and Springer (Thelwall & Kousha, 2015).

In most of the studies published hitherto, altmetrics impact has been measured based on counts of mentions (e.g., Haustein & Larivière, 2014), Mendeley reader counts (Mas-Bleda & Thelwall, 2016) or as field-normalized scores (e.g., Bornmann & Haunschild, 2016, 2018b; Fairclough & Thelwall, 2015; Haunschild & Bornmann, 2016). These measures can be used as an indication of attention to the publications. One of the problems with most altmetrics is currently that it is not clear what is measured (Bornmann & Haunschild, 2018a): is it only background noise or something substantial such as the importance of research for the health of many people? Does altmetrics measure actual impact, perfunctory attention, or broad popularity (Thelwall, Kousha, Dinsmore, & Dolby, 2016; Xia et al., 2016)? The results of Bornmann, Haunschild, and Adams (2018) show that social media counts do not measure the same societal impact as experts would do.

Instead of focusing on counts of mentions in social media – as analogous to citations to publications – we focus on comparing the topic networks in social media and in the scientific publications. We define topic networks as co-occurring hashtags on Twitter and author keywords in academic publications. Our research questions are:

> RQ 1: How do the scholarly networks of topics differ from the topics in public communication?
>
> RQ 2: To what extent are Twitter data useful for measuring the public discussion of scientific publications?

RQ 3: Are publications which are tweeted, not tweeted, mentioned in news media published in journals with a more general or a more special focus?

We compare four different networks and visualize the differences (and similarities) between the topics in academia, as represented by author keywords to the publications, and on social media, as represented by the hashtags attached to the tweets. We focus on Twitter data, because earlier studies point to the importance of Twitter as data for altmetrics, and because the data can be retrieved for large sets of publications. Contrary to most previous studies on Twitter metrics, we compare topical networks on Twitter and in the scientific publications. We show how Twitter data may be used for topic-related network analyses reflecting public discussions using our methodological approach. We compare the most frequently used publication venues for the publications of the four different networks.

# 2 Research on Twitter – a short overview

Twitter is a well-known social media platform for micro-blogging (i.e. allowing users to post short messages) which was founded in 2006 (https://www.twitter.com). Tweets are short messages with up to 280 characters in length (up to 140 characters until recently); if publications are mentioned, or a link to the publications is shared in tweets, the number of mentions or shared links can be counted for the use as altmetrics (Shema, Bar-Ilan, & Thelwall, 2014). The advantage of using Twitter data for measuring attention is that the attention usually happens immediately after appearance of a paper (days rather than years) (Wouters et al., 2015).

One of the problems with the usage of Twitter data for societal impact measurements is the restriction of tweets to only 280 (and previously 140) characters. This restriction results in tweeted texts with little content from which the reasons for tweeting can scarcely be deduced – in most of the cases (Haustein, Larivière, Thelwall, Amyot, & Peters, 2014; Taylor, 2013). “A typical tweet about a scientific article appears to be quite factual in its nature with little or no opinions expressed" (Vainio & Holmberg, 2017, p. 347). Friedrich, Bowman, and Haustein (2015) state similarly – based on their results – that "the majority of the processed tweets do not contain any sentiments and are therefore neither praise nor criticism but merely diffusion of the paper." Most of the tweets only repeat the title of a paper or a small part of its abstract (Thelwall & Kousha, 2015), or share a link to the publication. Furthermore, the meta-analysis published by Bornmann (2015) reveals that the correlation between Twitter counts and traditional citation counts is negligible. Thus, tweets do not measure what usually is considered as scholarly impact. It remains unknown whether tweets measure types of impact different from scholarly citations.

Recently, a specific kind of altmetrics (moving away from counting towards analyzing networks) has emerged that focuses on Twitter (Haustein, 2018; van Honk & Costas, 2016). Novel analyses of Twitter data aim at being used beyond counts of mentions and liking scores (which have the described problems in interpreting the meaningfulness of the results). Other foci can be explored when the data is analyzed as networks of actors on social media, or as networks of retweets, see for example Costas, de Rijcke, and Marres (2017) and Wouters, Zahedi, and Costas (2018). Recently, analysis of the number of re-tweets and hashtag couplings has been proposed (Costas, et al., 2017; van Honk & Costas, 2016). Analogously to citation networks, Twitter data can be used for analyzing coupling similar to bibliographic coupling or co-citations when several tweets mention the same academic paper (Costas, et al., 2017).

Twitter enables medium specific affordances, such as the use of hashtags that allow the users to search for tweets on specific topics. Several studies have focused on the use of hashtags. van Honk and Costas (2016), based on their analysis of the effects of hashtag use on the number of blog posts and Mendeley readers, conclude that the use of hashtags increases the public reception of academic articles. Haustein, Costas, and Lariviere (2015) compared social media metrics across scientific disciplines and concluded that overall every fifth article (21.5%) was tweeted at least once, yet Twitter (and other social media) are more frequently used for the social sciences, the humanities, biomedical, and health sciences than they are for natural sciences, engineering, mathematics, and computer sciences.

Recently, focus on network perspective as different from counting the number of mentions has been proposed for and applied to Twitter data. According to Robinson-García, van Leeuwen, and Ràfols (2017), an analysis of the data "in terms of networks (which can be more or less formal), can facilitate the understanding of the contexts (attributes of nodes), processes (links) and embedding (networks structure) of researchers" (p. 8). The authors analyzed productive interactions between academics and stakeholders from other sectors based on Twitter data. In a similar study, Hellsten and Leydesdorff (in press) analyzed Twitter data and mapped the co-occurrences of hashtags (as representation of topics) and usernames (as addressed actors). The resulting networks can show the relationships between three different types of nodes, i.e. authors, actors, and topics (Hellsten, Jacobs, & Wonneberger, 2019). The maps demonstrate how actors and topics are co-addressed in science-related communications. This new method (Hellsten & Leydesdorff, in press) opens opportunities for analysis of altmetrics data.

In this exploratory study, we take up the critique of using Twitter counts for measuring broad impact and follow the approaches by Robinson-García, et al. (2017) and Hellsten and Leydesdorff (in press) to use Twitter data for reflecting public discourses on scientific publications. According to Holmberg, Bowman, Haustein, and Peters (2014): "In addition to retweets and mentions, users also make use of the hashtag affordance to categorize, organize, and retrieve tweets. [...] As such, hashtags may resemble the traditional function of metadata by enhancing the description and retrievability of documents" (p. 3). In our opinion, these results suggest that hashtags on Twitter provide similarly useful information on publications as author keywords in scientific publications.

In the following, we compare (i) networks of author keywords in scientific publications that were tweeted with those that were not tweeted and (ii) author keyword networks versus networks of hashtags. We discuss whether these comparisons reveal public discussions about (climate change) research which can be analytically distinguished from research-focused discussions. Is the borderline between scholarly discourse and the public communication sharp or fuzzy? (iii) In order to see if we can find different characteristics about the main publication venues of the different types of publications, we compare the most frequent journals that published climate change research papers which were tweeted, not tweeted, or both tweeted and mentioned in news outlets.

# 3 Methods

## 3.1 Dataset used

We used the Web of Science (WoS, Clarivate Analytics) custom data of our in-house database derived from the Science Citation Index Expanded (SCI-E), Social Sciences Citation Index (SSCI), and Arts and Humanities Citation Index (AHCI) produced by Clarivate Analytics (Philadelphia, USA). A publication set containing most of the relevant literature regarding climate change research was compiled by Haunschild, Bornmann, and Marx (2016) using a sophisticated method known as “interactive query formulation” (Wacholder, 2011). In the first step, a set of key papers was retrieved. In the second step, the search query was reformulated according to the keyword analysis of the key papers. This procedure is repeated until most of the relevant publications are included in the results set. A detailed description of the search process for retrieving the relevant publications on climate change research can be found in Haunschild, et al. (2016). The search was restricted to the publication years 2011-2017 and to the document types “article” and “review”.

In total, the set of climate change publications consists of 176,122 papers of which 164,772 (93.6%) possess a digital object identifier (DOI) in the WoS database. We use the set of 164,772 papers to match them via the DOIs with our locally maintained database with data shared with us by the company Altmetric (see https://www.altmetric.com) on June 10, 2018.

The following information was thereafter appended to the DOIs: (1) links to the tweets which mentioned the corresponding paper, (2) number of Twitter accounts which mentioned the respective paper in a tweet, (3) respective number of tweets, and (4) number of mentions in news outlets of this same paper. Among the climate change papers with DOIs, 24.3% (n=39,985) of the climate change papers were mentioned on Twitter (in total 418,765 tweets), 16.5% (*n*=27,198) of the climate change papers were tweeted by at least two Twitter users (in total 404,227 tweets), and 3.5% (*n*=5,824) of the climate change papers were also mentioned in news outlets. For 403,918 tweets (99.9%), the URL was available in the Altmetric.com database.

## *3.2* Data extraction

We downloaded the 403,918 webpages of the tweets which mentioned a climate change paper. These webpages were downloaded using a dedicated routine written in Visual Basic (see http://leydesdorff.github.io/haunschild/index.html and https://www.leydesdorff.net/software/twitter for the dedicated routines with instructions). The routine Tweets1.exe downloads the webpages corresponding to the tweets in html format. In the case of 23,913 webpages, the requested URL did no longer exist. This usually happens when tweets are deleted. Therefore, we expect that 5.9% of the tweets were deleted and only the remaining 380,005 (94.1%) tweets could be used in our analysis. A routine Tweets1a.exe in the xBase language, but compiled with Harbour under Linux (see https://github.com/harbour/core) parses the downloaded webpages to extract the tweets, authors, time, and year. This information is stored in a database file (*tweets.dbf*). Another routine (Year.exe) was used to extract the tweets from the years 2011-2017. 15,659 tweets were tweeted after 2017. They were removed from our dataset. The remaining 364,346 tweets were analyzed in the following.

We chose the time period of 2011 to 2017, because Altmetric.com started to cover Twitter in 2011 and our publication set does not contain papers published after 2017. Further routines (frqtwt.exe) and (tweet.exe) were used for producing a ranked frequency distribution of terms and a Pajek file corresponding to a word/document matrix. Among other things tweet.exe produces a cosine-normalized (McGill, 1983) term co-occurrence matrix (see https://www.leydesdorff.net/software/twitter). The routines are available from http://leydesdorff.github.io/haunschild/index.html (source codes on request).

### 3.3 Hashtags

We are interested in all hashtags (terms starting with the # sign) including name variants. No stop-word list was needed because all hashtags are meaningful. In total, we found 236,696 hashtags in 364,346 tweets which referred to 27,198 papers. Therefore, we have 0.65 hashtags per tweet, 8.7 hashtags per tweeted paper, and 1.44 hashtags per paper with DOI (tweeted or not). 127,839 tweets (35.1%) contained at least one hashtag. Figure 1 shows that the distribution of hashtags per paper is rather skewed. The paper tweeted with most hashtags has the following DOI: 10.1016/j.rser.2017.01.181. This paper has been tweeted 869 times with 4266 hashtags (4.6 hashtags per tweet on average). However, it occurs much more frequently that papers are tweeted with no or only a single hashtag.

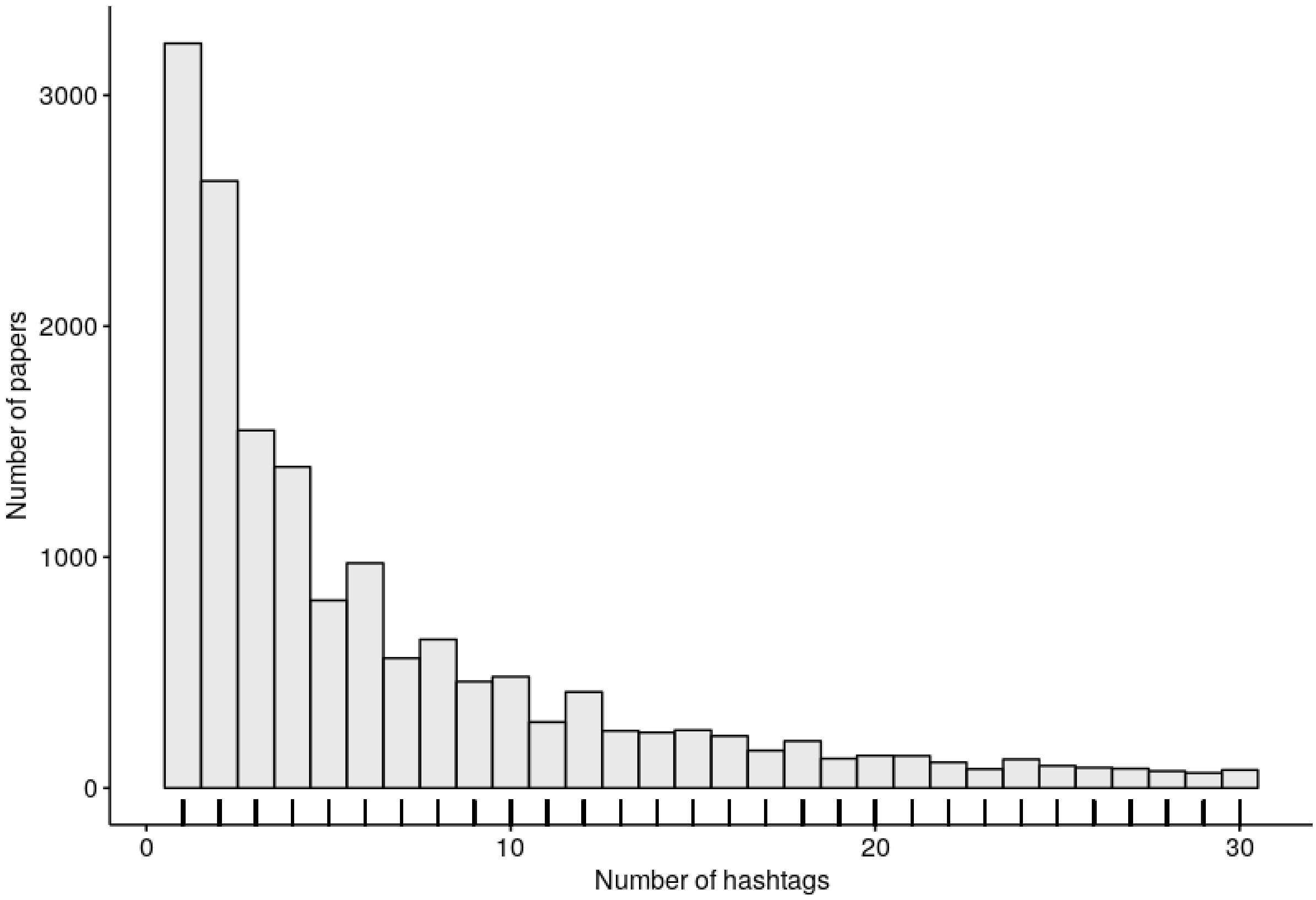

**Figure 1:** Frequency distribution of hashtags per paper for the region from 1 to 30 hashtags per paper.

The most frequent hashtags were determined from the output of the routine frqtwt.exe and chosen for further analysis (see below). Within the most frequently occurring hashtags, some clear synonyms appeared. We merged the hashtags (i) “#ANTARCTIC” and “#ANTARCTICA” into “#ANTARCTICA”, (ii) “#BIODIVERSIDAD” and “#BIODIVERSITY” into “#BIODIVERSITY”, (iii) “#CAMBIOCLIMATICO” and “#CAMBIOCLIMTICO” into “#CAMBIOCLIMATICO”, (iv) “#COP21” and “#COP22” into “#COP” (“Conference of the Parties” to the “United Nations Framework Convention on Climate Change”), (v) “#FOREST” and “#FORESTS” into “#FORESTS”, (vi) “#OA” and “#OPENACCESS” into “#OPENACCESS”, and (vii) “#EXTINCIÓN”, “#EXTINCTION”, and “#EXTINCTIONS” into “#EXTINCTION”.

### 3.4. Author keywords

For a comparison of the hashtags with keywords, we exported the author keywords of the climate change papers published between 2011 and 2017. 75.9% of the climate change papers (n = 125,003) with a DOI had also at least one author keyword attached to them. Whitespaces and hyphens within keywords were replaced with underscores so that they can all be treated as word occurrence. A keyword frequency list was exported from our bibliometric database. In a separate export, all author keywords which belong to a single paper were grouped in a line of the file *text.txt* to analyze the keywords analogously to the hashtags with the routine tweet.exe.

We exported four different sets of author keywords: (1) author keywords of papers tweeted by at least two Twitter accounts and mentioned in news outlets at least once, (2) author keywords of papers tweeted by at least two Twitter accounts, (3) author keywords of not tweeted papers, and (4) author keywords of all climate change publications. We aimed to use the most frequently occurring (top-1%) author keywords of the first set. 9,236 different author keywords appear for the set of 5,824 papers tweeted by at least two accounts and mentioned at least once in news outlets. Author keywords around the rank number 92 are tied in their number of occurrences (n=13). Therefore, we use author keywords which appeared more than 13 times for this set. These are the top-85 author keywords of this set. Within these top-85 author keywords, no clear synonyms appeared. We aimed to use a similar number of top author keywords for the other sets in order to compare networks of the same size.

The second set of top-85 author keywords (from papers which were tweeted by at least two Twitter accounts) contained some obvious synonyms. We merged the author keywords (i) “greenhouse_gas” and “greenhouse_gases” into “greenhouse_gases”, (ii) “modeling”, “modelling”, and “models_and_modeling” into “modeling”, and (iii) “palaeoclimate” and “paleoclimate” into “paleoclimate”. Additional lower ranked author keywords were included.

However, in total only 84 author keywords could be included because the next author keyword is tied with three other author keywords.

The third set of top-85 author keywords (from papers which were not tweeted) also contained some synonyms. We merged the author keywords (i) “carbon_dioxide” and “co2” into “carbon_dioxide”, (ii) “greenhouse_gas” and “greenhouse_gases” into “greenhouse_gases”, (iii) “lca”, “life_cycle_assessment_(lca)” and “life_cycle_assessment” into “life_cycle_assessment”, (iv) “modeling” and “modelling” into “modeling” and (v) “palaeoclimate” und “paleoclimate” into “paleoclimate”. Additional lower ranked author keywords were included to reach a set of top-85 author keywords.

Finally, the fourth set of top-85 author keywords (from all papers) contained the same synonyms as the third set.

### 3.5 Visualization

The resulting files (containing cosine-normalized distributions of terms in the Pajek format, see http://mrvar.fdv.uni-lj.si/pajek) were laid-out using the algorithm of Kamada and Kawai (1989) in Pajek and then exported to VOSviewer 1.6.10 (see http://www.vosviewer.com) for visualizations. The clustering algorithm in VOSviewer was employed with 10 random starts, 10 iterations, a random seed of 0, and the option “merge small clusters” enabled. The resolution parameter and minimum cluster size were chosen to obtain four clusters. These two values are reported in the figure captions. The size of a node indicates its frequency of co-occurrence with all terms on the map. Lines between two nodes and their thickness indicate the co-occurrence frequency of these specific terms.

### 3.6 Journals

As another indicator of the focus of papers which were tweeted, papers which were tweeted and were mentioned in the news, papers which were not tweeted, and all papers, we compare the sets of top-20 journals in which these papers are published. Each top-20 set of journals has a significant overlap with other top-20 sets but some journals are unique to specific sets. The super-set of all four top-20 journal sets consists of 34 journals.

# 4 Results

## 4.1. Author keywords

In the first step, we discuss the results of co-occurring author keywords, and compare across publications that were tweeted, those not tweeted, and those that were both tweeted and mentioned in the news media. Figure 2 shows the semantic map of the top-85 author keywords of climate change publications during the time period 2011-2017. The red cluster contains various important author keywords of climate change publications. The green cluster is focused on environmental aspects. The blue cluster contains mostly keywords related to adaptation to climate change. The yellow cluster contains terms related to policy topics and climate change mitigation. Mainly, the expected keywords for the overall climate change literature in this period are visible. In addition to the search terms and their synonyms (climate, climate change, global warming) the following keywords and topics appear: generic terms related to the greenhouse effect ("greenhouse_gases" and "greenhouse_gas_emissions"), more specific terms related to the greenhouse effect ("Carbon", "Carbon_dioxide", "Methane", and "Nitrous_oxide"), the most affected countries and regions ("Australia", "China", "India", "Europe", and "Arctic"), geography and ecosystem-related terms ("Biogeography", "Phylogeography", "Permafrost", "Sea_ice", "Sea_surface_temperature"), specific impacts ("Drought", "Fire", "Precipitation", "Rainfall", "Water_quality", and "Soil_moisture"), most important keywords with regard to the past climate ("Dendrochronology", "Holocene", "Paleoclimate", "Pollen"), and to the future climate ("Climate_models", "Modeling" and "Uncertainty"); also, the expected policy-relevant terms appear ("Climate_policy", "Adaptation", "Mitigation", "Conservation",

“Vulnerability”, “Renewable_energy“, “Energy_efficiency“, and “Sustainable_development“).

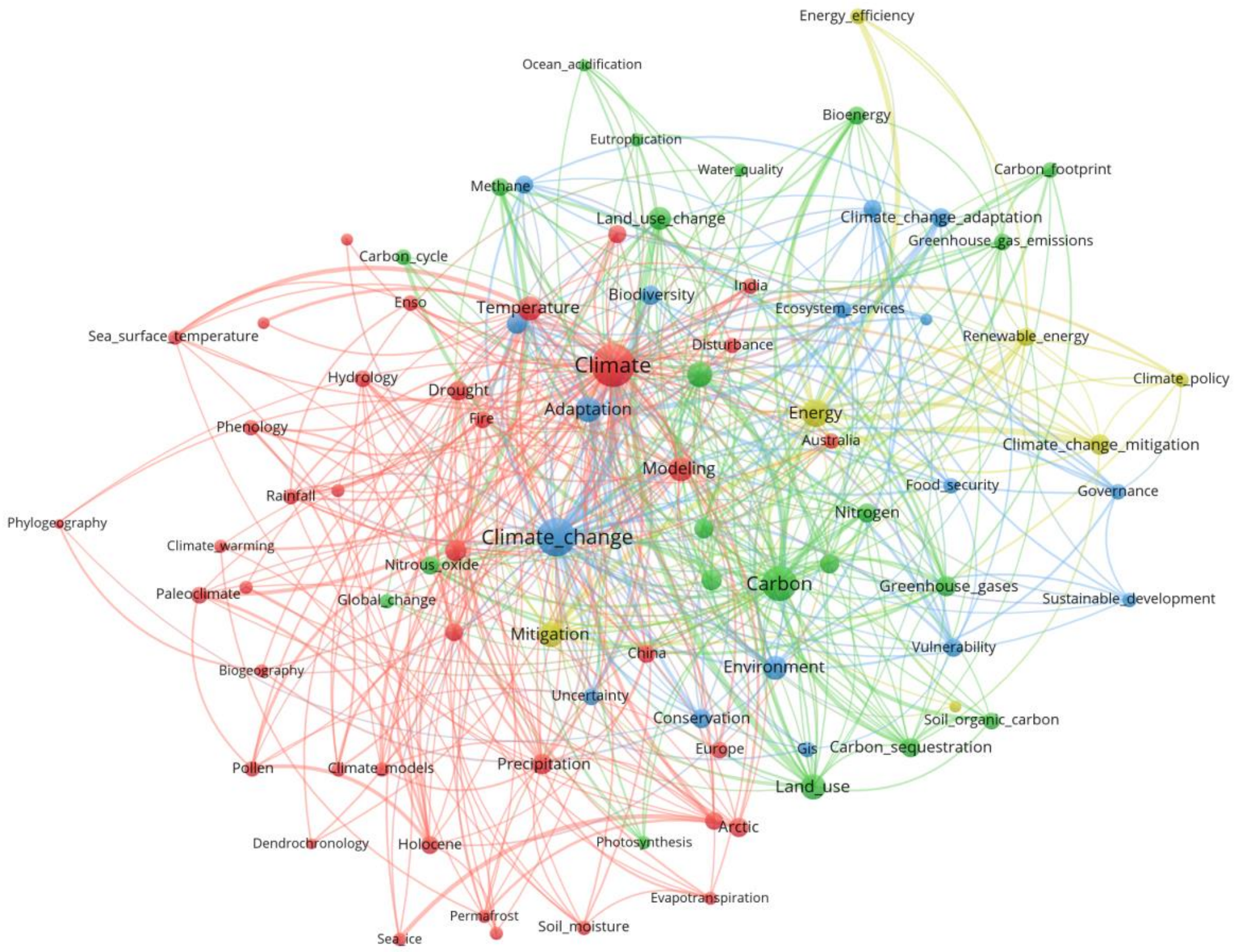

**Figure 2:** Top-85 author keywords of climate change research papers published between 2011 and 2017. A resolution parameter of 0.94 with a minimum cluster size of 1 was used. An interactive version of this network can be viewed at http://tinyurl.com/y2tuz9va. Note that the color scheme may be different in the interactive version.

Figure 3 shows the semantic map of the top-85 keywords of not tweeted papers for comparison. Overall, the corpus of author keywords is very similar in Figure 2 and Figure 3, although cluster assignments have sometimes changed. This is due to the sensitivity of the clustering algorithm to minor changes of the text corpus.

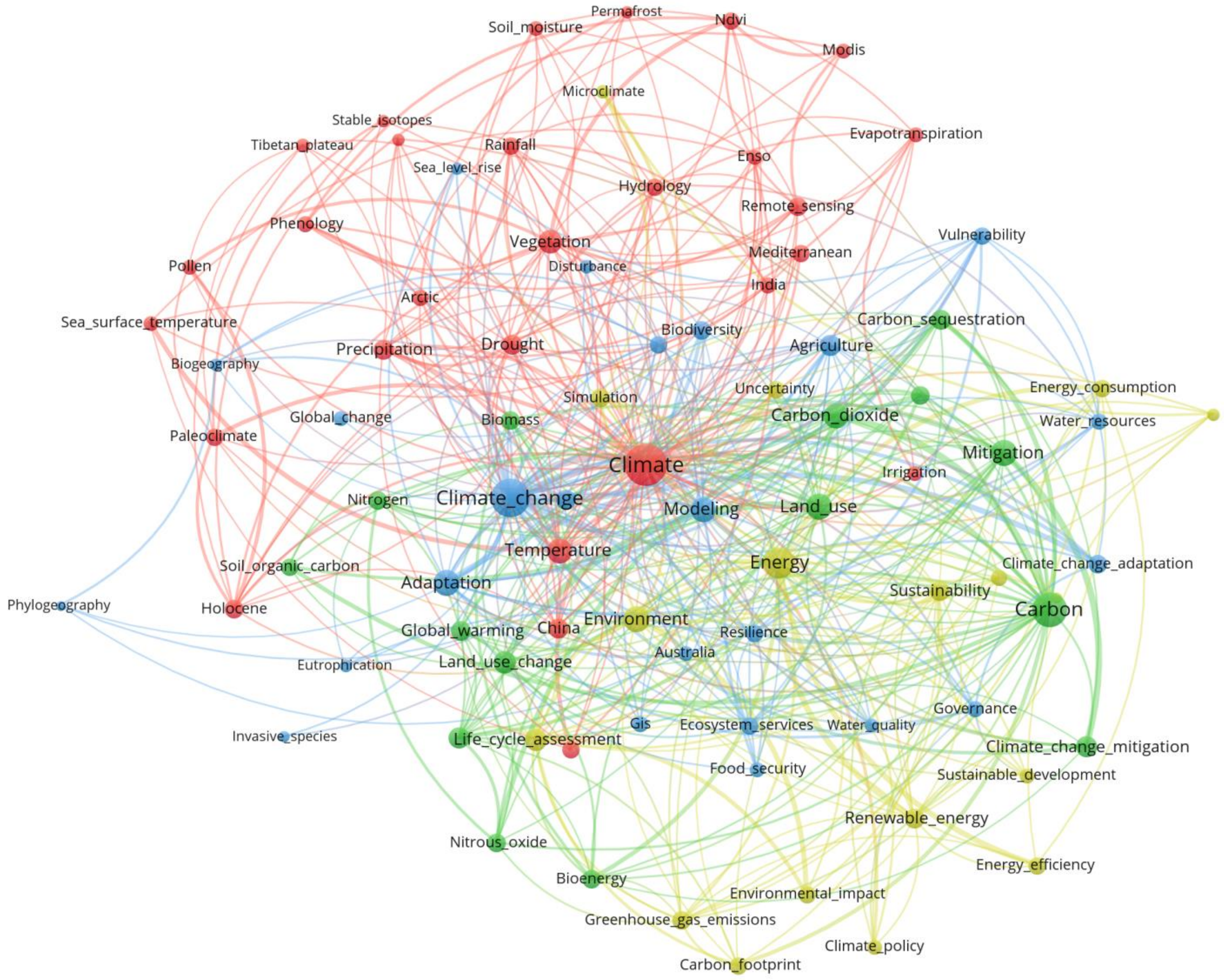

**Figure 3:** Top-85 author keywords of not tweeted climate change research papers published between 2011 and 2017. A resolution parameter of 0.75 with a minimum cluster size of 1 was used. An interactive version of this network can be viewed at http://tinyurl.com/y4kkgade. Note that the color scheme may be different in the interactive version.

Figure 4 shows the network of the top-84 keywords of climate change papers which were tweeted by at least two Twitter accounts. In contrast to Figure 2 and Figure 3, the policy-relevant terms of the yellow cluster in Figure 4 appear more numerous, whereas the basic climate change research terms of the red cluster have become reduced. However, the keywords related to climate modelling appear more pronounced (“Cmip5”, “Climate_models”). The keyword “Cmip5” stands for “Coupled Model Intercomparison Project Phase 5”. The term “Management” appears as a new pronounced term within the blue cluster. The new keyword "Redd" is short for "Reducing Emissions from Deforestation and Forest Degradation".

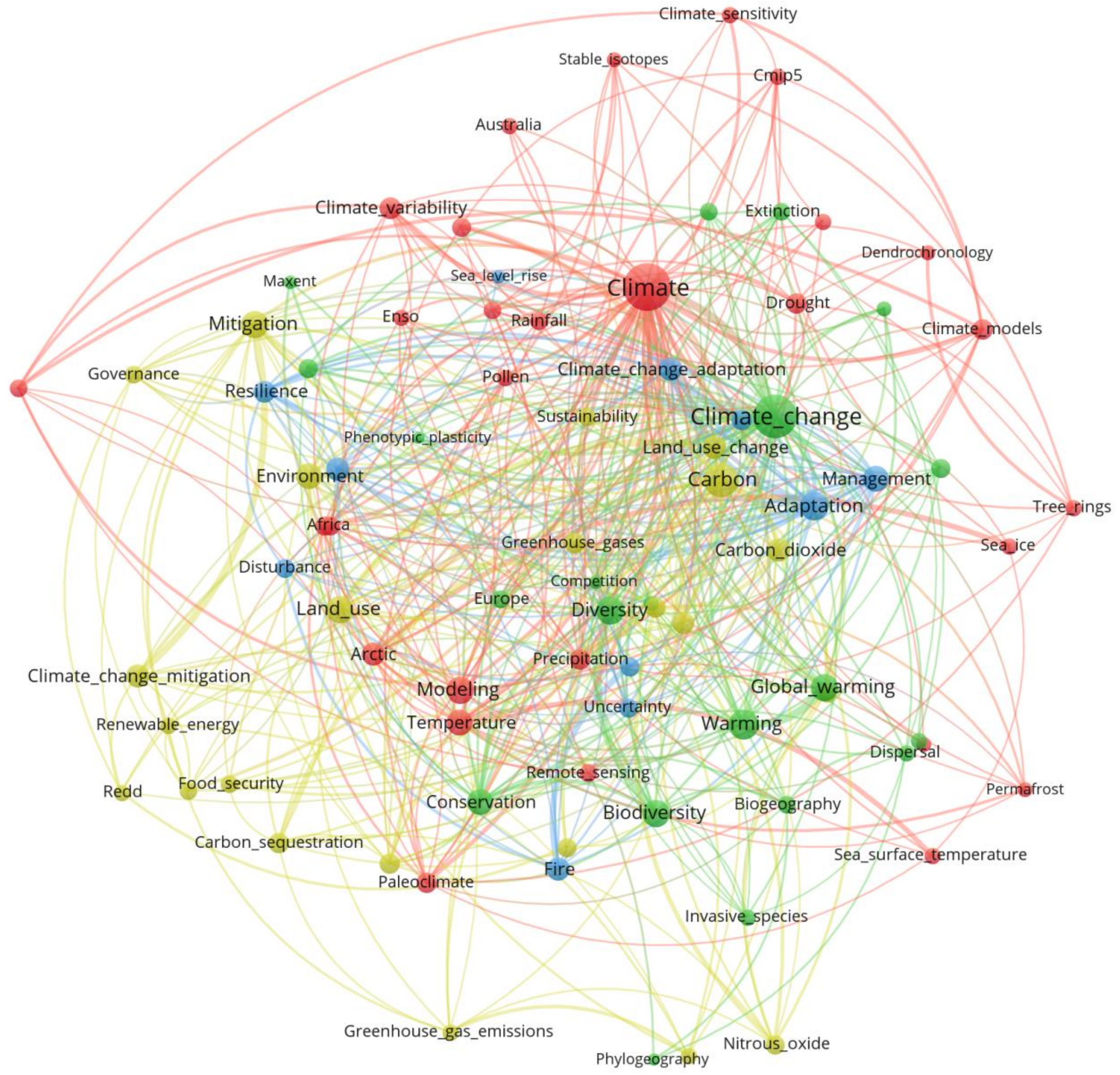

**Figure 4:** Top-84 author keywords of climate change research papers published between 2011 and 2017 and mentioned by at least two Twitter accounts. A resolution parameter of 0.70 with a minimum cluster size of 1 was used. An interactive version of this network can be viewed at http://tinyurl.com/y3avslu6. Note that the color scheme may be different in the interactive version.

Figure 5 shows the network of the top-85 author keywords of climate change research papers published between 2011 and 2017 which were mentioned by at least two Twitter accounts and at least once by a news outlet. We used mentions in news outlets as a second criterion (in addition to multiple mentions of papers on Twitter), as we expected more focused networks of public discussions on climate change. In most of the cases, only those papers are selected for news reports which are of interest for a broader audience. As is to be expected, the scope of the keywords in Figure 4 is broader than in Figure 5.

The semantic map in Figure 5 shows three larger clusters in red, blue, and yellow as well as one smaller cluster in green. The red cluster contains the basic climate change research relevant keywords. The blue cluster containing the environment related keywords has been expanded on the expense of the green cluster containing biology related terms. Among others, “Alaska”, Coral_reefs”, “Policy”, “Public_health”, and “Public_opinion” appear as new terms.

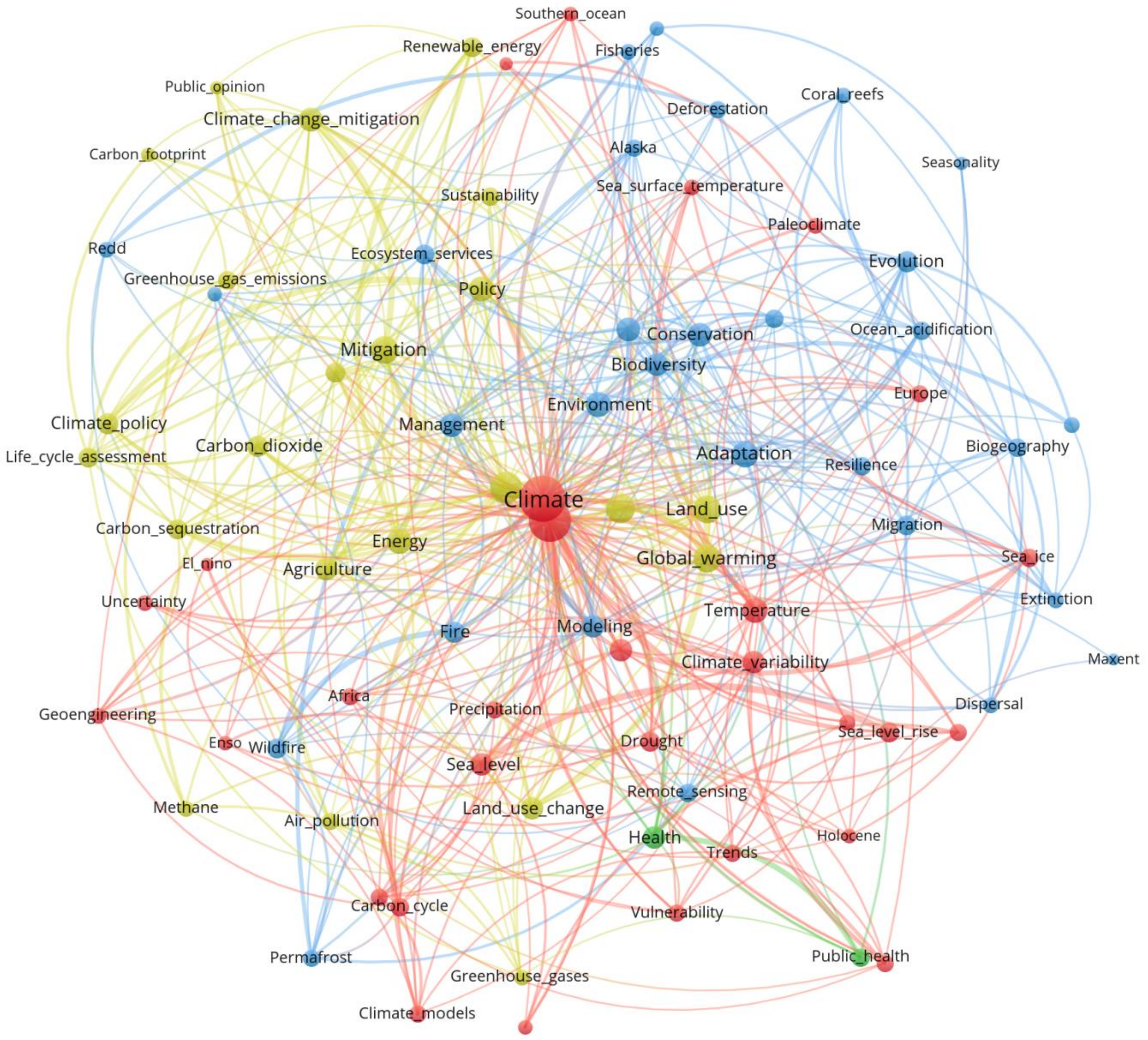

**Figure 5:** Top-85 author keywords of climate change research papers published between 2011 and 2017, mentioned by at least two Twitter accounts, and mentioned by a news outlet. A resolution parameter of 0.70 with a minimum cluster size of 1 was used. An interactive version of this network can be viewed at http://tinyurl.com/y2ezyxwu. Note that the color scheme may be different in the interactive version.

Overall, the topics covered by Figure 4 and Figure 5 are again similar but the topics are more focused towards environment-related issues in Figure 5 due to the restriction of papers which also were mentioned in news outlets. The author keyword networks in Figure 2 and Figure 3 (semantic maps of all and not tweeted papers) do not show a special focus within climate change research as the author keyword networks in Figure 4 and Figure 5 (semantic maps of tweeted papers) do. Some more professional keywords (scientific jargon) appear only in the semantic maps of Figure 2 and Figure 3 or remain below the threshold of the top-85 lists in the semantic maps of Figure 4 and Figure 5; for example, "Eutrophication", "Modis" (moderate-resolution imaging spectroradiometer), and "Evapotranspiration". Table 1 shows the overlap between the top-85 author keywords.

**Table 1:** Overlap between top-85 author keywords. The lower triangle shows the absolute number of overlapping keywords and the upper triangle shows the proportion of overlapping keywords.

| | *All* | *Not tweeted* | *Tweeted* | *Tweeted and mentioned in the news* |
|---|---|---|---|---|
| *All* | 85 | 88.2% | 77.4% | 65.9% |
| *Not tweeted* | 75 | 85 | 69.0% | 57.6% |
| *Tweeted* | 65 | 58 | 84 | 75.0% |
| *Tweeted and mentioned in the news* | 56 | 49 | 63 | 85 |

The commonalities of the top-85 author keywords decrease in the following order: all author keywords > author keywords of not tweeted papers > author keywords of papers mentioned by at least two Twitter accounts > author keywords of papers mentioned by at least two Twitter accounts and mentioned by news outlets.

In sum, our results regarding author keywords indicate that publications using scientific jargon are less likely to be tweeted than publications using more general keywords. The general public seems to be more interested in climate forecast and consequences of climate change to the environment and to adaptation, mitigation and management issues rather than in the methodology of climate change research and causes of climate change. The not tweeted papers accentuate the specific greenhouse gases (carbon dioxide, nitrous oxide, and methane) and the phenomena caused by climate change (drought, hydrology, precipitation, and temperature). The tweeted papers have a stronger focus on modelling the future climate. With regard to the tweeted papers and mentioned in news, climate policy related issues (e.g., adaption, mitigation, management, and energy) appear more accentuated. Overall, we can say that papers related to the impact of climate change on the biosphere (flora, fauna, agriculture, and health) generated a large amount of public impact.

## 4.2 Hashtags

In the second part, we focus on the co-occurring hashtags. Figure 6 shows the network of the top-85 hashtags from tweets between 2011 and 2017 which mentioned a climate change research paper. The network of hashtags shows many hashtags related to flora and fauna (e.g., "#BIRDS", "#ORNITHOLOGY", "#FISHERIES", "#AGRICULTURE", "#FORESTS", "#WILDFIRE". Furthermore, the semantic map of hashtags shows some journal or publisher names as hashtags; for example, #PLOS, #NATURE, and #SCIENCE. However, manual checks revealed that most of the times #NATURE and #SCIENCE refer to nature and science (not the journals). When Twitter users wish to refer to the journals *Nature* and *Science* they typically use the twitter handles @Nature and @Science. This is obviously different in the use of #PLOS, #PLOSBIOLOGY, and #PLOSONE.

Overall the semantic map of hashtags (Figure 6) has a strong focus on flora, fauna, and adaptation to climate change and does not represent climate change research as a whole. This result is in line with previous research on Twitter use in the context of climate change, in particular tweets about the 2013 publication of the Intergovernmental Panel on Climate Change (IPCC) where the focus is on food, and agriculture was one of the core topics on Twitter (Pearce, Holmberg, Hellsten, & Nerlich, 2014). None of the top-85 hashtags is related to climate change mitigation. Mitigation (as it does not appear within the top-85 hashtags) seems to play a minor role for twitter users in comparison to adaptation (which is one of the medium-sized) nodes in Figure 6. Climate change impacts on the biosphere appear more specific concerning flora and fauna (birds, ornithology, coral, fish, and forests) compared to the author keywords of tweeted papers (Figure 4) and some most affected and sensible regions (Arctic, Antarctica, and China). The more specific hashtags may indicate deeper engagement of the Twitter user with the publication than in the case of hashtags which also appear as author keywords.

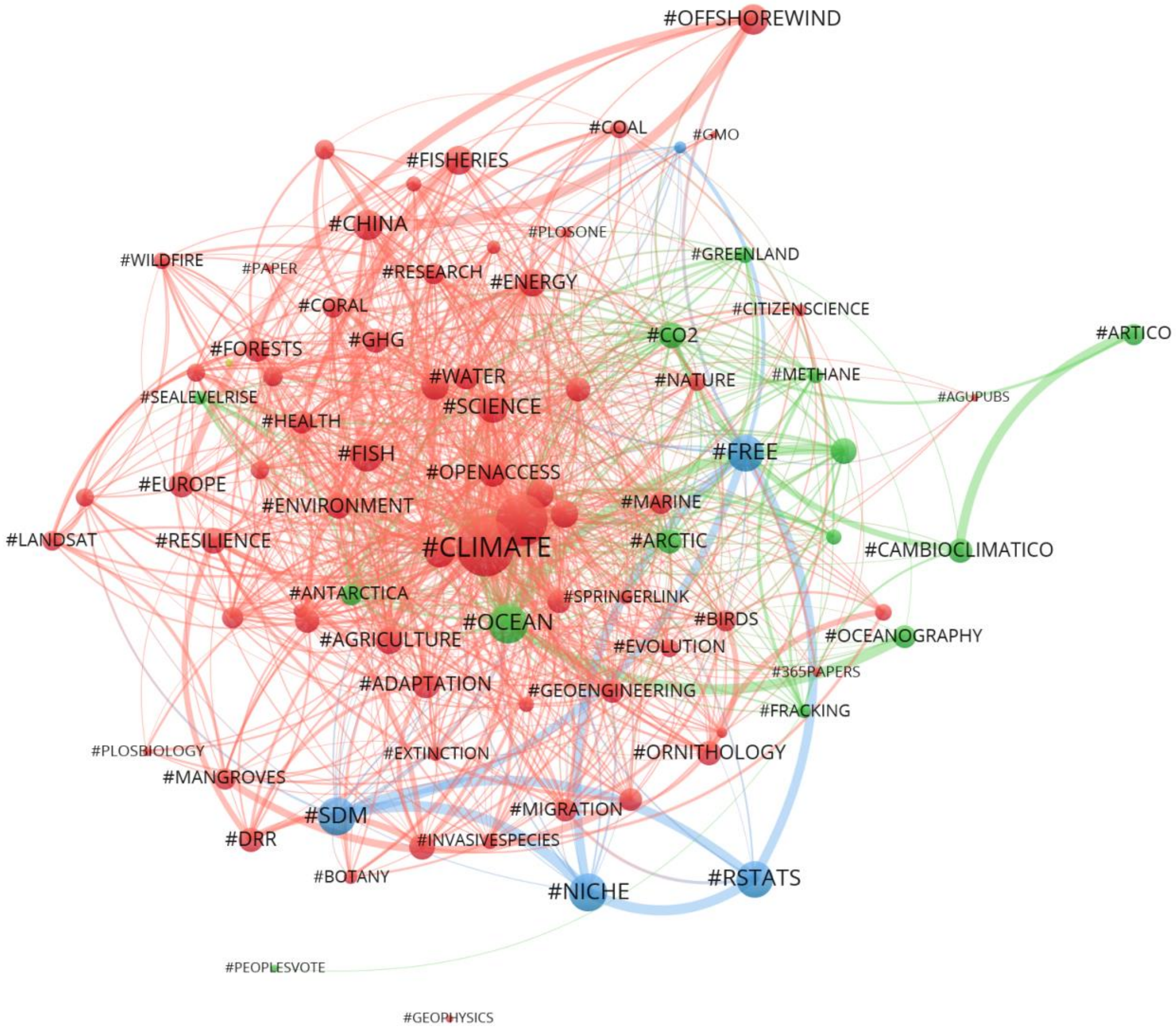

**Figure 6:** Top-85 hashtags from tweets between 2011 and 2017 which mentioned a climate change research paper. A resolution parameter of 0.50 with a minimum cluster size of 5 was used. An interactive version of this network can be viewed at http://tinyurl.com/y5red2lf. Note that the color scheme may be different in the interactive version.

### 4.3 Journals

In the final step, we show the results of the journals in which the papers were published and compare the scope of the journals across the different types (tweeted, not tweeted, ...). Figure 7 shows the relative importance of the top journals for the four different sets of publications: (i) all climate change papers (red bars), (ii) not tweeted climate change papers (green bars), (iii) climate change papers which were mentioned by at least two Twitter accounts (blue bars), and (iv) climate change papers which were mentioned by at least two Twitter accounts and at least one news outlet (pink bars). Many journals appear in all four top-20 lists but with different importance, e.g., journals such as *PLoS One* and *Climatic Change* are much more important for papers which are tweeted (and mentioned in the news) than for not tweeted papers or all climate change papers. None of the journals in the top-20 lists shows an equal importance for all four sets, although *Quaternary Science Reviews* is rather close. Other journals, e.g., *Nature* and *Science* (i.e., journals with more general content) appear only in the top-20 list of papers which were mentioned by at least two Twitter accounts and at least one news outlet. *PNAS* appears in both sets which contain tweeted papers. The three journals *Nature*, *Science*, and *PNAS* have a very high reputation in science. They contain shorter contributions and have a multidisciplinary scope. Papers in these journals seem to be of high interest for the general public. It seems likely that this relates at least partly to the prestige of these journals, their visibility among journalists, and the efforts they make to generate publicity for the articles they publish. More specialized journals such as, e.g., *Applied Energy*, *Journal of Hydrology*, and *Theoretical and Applied Climatology* appear only in the top-20 lists of all and/or not tweeted papers of climate change research. In total, the analysis of the top-20 journals of the four different types of climate change publications substantiates our results from the comparison of author keywords. The general public is mainly interested in more general research results and less interested in the details of climate change research.

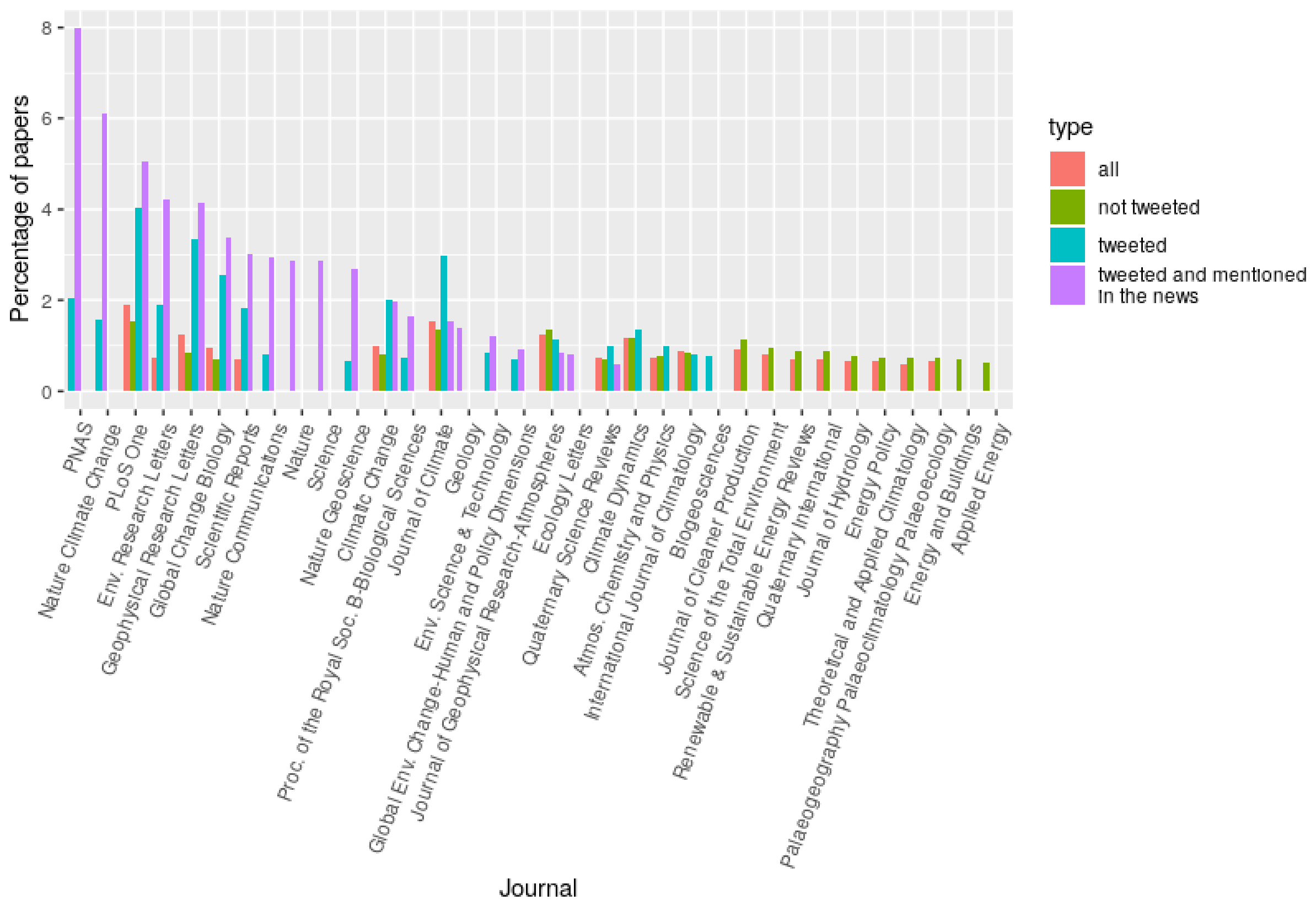

**Figure 7:** Most frequent journals of climate change research papers ordered by the relative importance for the papers which were mentioned by at least two Twitter accounts and appeared at least once in news media.

# 5 Discussion

Hashtags can be considered as meta-information regarding tweets similarly to author keywords were considered as meta-information regarding scientific publications. Our exploratory study is based on a comprehensive publication set of papers about climate change research. We are interested in whether Twitter data are able to reveal topics of public discussions which can be compared to research-focused topics. This provides useful information about which scientific publications enter the public discussion on Twitter, i.e., which publications may have broader societal impact beyond academia.

Twitter counts have already been used in many scientometrics studies, but the meaningfulness of the data for societal impact measurements in research evaluations has been questioned. We proposed to focus on the content of tweets and scientific publications by analyzing the hashtags in tweets and the author keywords in scientific publications. We used a recently developed network approach in which we analyzed the topics of tweeted publications and compared them with the topics of scientific publications which were not tweeted. In other words, we contrasted publications that were tweeted with those that were not tweeted.

The not tweeted papers accentuate the specific greenhouse gases (carbon dioxide, nitrous oxide, and methane) and the phenomena caused by climate change (drought, hydrology, precipitation, and temperature). The tweeted papers have a stronger focus on modelling the future climate. With regard to the tweeted papers and mentioned in news, climate policy related issues (e.g., adaption, mitigation, management, and energy) appear more accentuated. In the case of hashtags, climate change impacts on the biosphere appear more specific concerning flora and fauna (birds, ornithology, coral, fish, and forests) and some most affected and sensible regions (Arctic, Antarctica, and China). Overall, we can say that papers related to the impact of climate change on the biosphere (flora, fauna, agriculture, and health) generated a large amount of public (societal) impact. Note that we understand impact as public discussions of papers (or topics) but not practical realizations of climate change research outcomes in society. The results of Bornmann, et al. (2018) point out that experts are more interested in these practical realizations which are not reflected, however, in altmetrics impact data (which seem to reflect public discussions).

Our results indicate that the most tweeted topics regarding climate change research are its consequences to humans. Adaptation to climate change is more important in the hashtags than mitigation of climate change which is in contrast to climate change in general where both topics are about equally relevant. Twitter users are especially likely to tweet about climate change publications which forecast effects of a changing climate on the environment. Policy-related issues become more important in the set of tweeted publications. This focus of Twitter users is reflected in both the author keywords of the tweeted papers and the hashtags the Twitter users choose themselves. In contrast to the author keywords of all and the not twittered publications, the hashtags do not mirror a scientific discourse; neither with regard to the understanding of the climate system (the primary aim of basic research) nor concerning the evolution of climate change from a hypothesis to a widely accepted fact (at least within the scientific community). This answers our first two research questions.

The comparison of the networks based on publications which were tweeted and mentioned in news outlets with those publications that were not tweeted reveals the public discussion around climate change topics which can be separated from topics being of interest for academia only. Overall, our results indicate that publications using scientific jargon are less likely to be tweeted than publications using more general keywords. This result is substantiated by a comparison of most frequently appearing journals in the set of tweeted or not tweeted publications. More general scientific journals like *Nature* and *Science* have a much higher weight for publications which were tweeted and mentioned in the news, while journals with a more specific focus like *Applied Energy*, *Journal of Hydrology*, and *Theoretical and Applied Climatology* have a much higher importance for not tweeted publications or all climate change publications. This is in line with results from Leydesdorff and Zhou (2007) who found that patents cited publications in the journals *Nature* and *Science* very frequently. The general public seems to be interested in shorter contributions of inter-, cross-, and multi-disciplinary climate change research. This answers our third research question. Future work could study if this is also the case for other topics.

In this exploratory study, we used climate change as an example to demonstrate a new approach of meaningful analysis of Twitter data (beyond analyzing Twitter counts – as comparable to citations). Since the tools for undertaking the necessary analyses steps are publicly available, our approach can be used for other datasets as well.

# 6 Limitations

The main limitation of this exploratory study is that discussions derived from tweeting papers are not captured by Altmetric.com, i.e., are not connected to the publications as the link to the publications is usually not repeated in every part of the discussion. Capturing such discussion is part of future work.

Another limitation of this study is that it is focused on climate change research. Similar studies with a focus on other topics should be performed because there are disciplinary differences in Twitter usage: “For instance, conversations in tweets in one small study were more common in Digital Humanities and Cognitive Science (both 38%), Astrophysics (31%) and History of Science than in Biochemistry and Economics (both 16%). In Biochemistry, 42% of tweets are retweets, whereas in nine other fields the proportion varied from 18% in Social Network Analysis to 33% in Sociology” (Thelwall & Kousha, 2015, p. 612). This approach can further be refined to provide guidelines on which kind of publications are more likely to be tweeted about and hence might have the potential for broader social impact also beyond the sciences.

Some scientometrics studies have pointed to the problem of background noise in Twitter data, e.g.: tweets from the author of a paper or the journal where it is published in. In this study, we proposed a two-step method to reduce noise in the analysis of Twitter data and focus on the tweets by the public: First, we considered only publications mentioned by at least two, separate Twitter accounts. Second, we removed even more noise by focusing on papers not only mentioned by at least two Twitter accounts but also mentioned in news outlets. The first step is already recommended by Altmetric.com. As also scientists are active on Twitter, the second step (to require also a mention in the news) is important to obtain a focus on the general public. We encourage other scientometricians working with Twitter data to use this approach (or similar approaches) to receive more meaningful results focused on the general public based on Twitter data. We made our software available for this purpose at http://leydesdorff.github.io/haunschild/index.html. Another possible approach proposed previously is to reduce noise in Twitter data by removing self-tweets (Sankar, 2015). However, this does not capture semi-automatic tweets from journal accounts.

Finally, another limitation is our focus on the most frequently used author keywords, hashtags, and publishing venues. More focused research topic, for example, a specific sub-field of climate change research, would allow for analyzing the full data set.

## Acknowledgements

The bibliometric data used in this paper are from an in-house database developed and maintained by the Max Planck Digital Library (MPDL, Munich) and derived from the Science Citation Index Expanded (SCI-E), Social Sciences Citation Index (SSCI), Arts and Humanities Citation Index (AHCI) prepared by Clarivate Analytics, formerly the IP & Science business of Clarivate Analytics (Philadelphia, Pennsylvania, USA). The twitter and news data were retrieved from our locally maintained database with data shared with some of us (RH and LB) by the company Altmetric on June 10, 2018.

# References

Bornmann, L. (2014). Do altmetrics point to the broader impact of research? An overview of benefits and disadvantages of altmetrics. *Journal of Informetrics, 8*(4), 895-903. doi: 10.1016/j.joi.2014.09.005.

Bornmann, L. (2015). Alternative metrics in scientometrics: A meta-analysis of research into three altmetrics. *Scientometrics, 103*(3), 1123-1144.

Bornmann, L. (2016). Scientific Revolution in Scientometrics: The Broadening of Impact from Citation to Societal. In C. R. Sugimoto (Ed.), *Theories of informetrics and scholarly communication* (pp. 347-359). Berlin, Germany: De Gruyter.

Bornmann, L., & Haunschild, R. (2016). How to normalize Twitter counts? A first attempt based on journals in the Twitter Index. *Scientometrics, 107*(3), 1405-1422. doi: 10.1007/s11192-016-1893-6.

Bornmann, L., & Haunschild, R. (2017). Does evaluative scientometrics lose its main focus on scientific quality by the new orientation towards societal impact? *Scientometrics, 110*(2), 937-943. doi: 10.1007/s11192-016-2200-2.

Bornmann, L., & Haunschild, R. (2018a). Do altmetrics correlate with the quality of papers? A large-scale empirical study based on F1000Prime data. *PLOS ONE, 13*(5), e0197133. doi: 10.1371/journal.pone.0197133.

Bornmann, L., & Haunschild, R. (2018b). Normalization of zero-inflated data: An empirical analysis of a new indicator family and its use with altmetrics data. *Journal of Informetrics, 12*(3), 998-1011. doi: 10.1016/j.joi.2018.01.010.

Bornmann, L., Haunschild, R., & Adams, J. (2018). Do altmetrics assess societal impact in the same way as case studies? An empirical analysis testing the convergent validity of altmetrics based on data from the UK Research Excellence Framework (REF). Retrieved August 25, 2018, from https://arxiv.org/abs/1807.03977

Costas, R., de Rijcke, S., & Marres, N. (2017). *Beyond the dependencies of altmetrics: Conceptualizing 'heterogenous couplings' between social media and science*. Paper presented at the Altmetrics 17, Toronto. www.altmetrics.org/wp-content/uploads/2017/09/altmetrics17_paper_4.pdf

Fairclough, R., & Thelwall, M. (2015). National research impact indicators from Mendeley readers. *Journal of Informetrics, 9*(4), 845-859. doi: http://dx.doi.org/10.1016/j.joi.2015.08.003.

Friedrich, N., Bowman, T. D., & Haustein, S. (2015). Do tweets to scientific articles contain positive or negative sentiments? Retrieved January 28, 2016

González-Valiente, C. L., Pacheco-Mendoza, J., & Arencibia-Jorge, R. (2016). A review of altmetrics as an emerging discipline for research evaluation. *Learned Publishing, 29*(4), 229-238. doi: 10.1002/leap.1043.

Haunschild, R., & Bornmann, L. (2016). Normalization of Mendeley reader counts for impact assessment. *Journal of Informetrics, 10*(1), 62-73. doi: 10.1016/j.joi.2015.11.003.

Haunschild, R., Bornmann, L., & Marx, W. (2016). Climate Change Research in View of Bibliometrics. *Plos One, 11*(7), 19. doi: 10.1371/journal.pone.0160393.

Haustein, S. (2018). Scholarly Twitter metrics. *eprint arXiv:1806.02201*, arXiv:1806.02201.

Haustein, S., Costas, R., & Lariviere, V. (2015). Characterizing Social Media Metrics of Scholarly Papers: The Effect of Document Properties and Collaboration Patterns. *Plos One, 10*(3). doi: 10.1371/journal.pone.0120495.

Haustein, S., & Larivière, V. (2014). *Mendeley as a Source of Readership by Students and Postdocs? Evaluating Article Usage by Academic Status.* Paper presented at the Proceedings of the IATUL Conferences. Paper 2.

Haustein, S., Larivière, V., Thelwall, M., Amyot, D., & Peters, I. (2014). Tweets vs. Mendeley readers: How do these two social media metrics differ? *Information Technology, 56*(5), 207-215.
Hellsten, I., Jacobs, S., & Wonneberger, A. (2019). Active and passive stakeholders in issue arenas: A communication network approach to the bird flu debate on Twitter. *Public Relations Review, 45*(1), 35-48. doi: https://doi.org/10.1016/j.pubrev.2018.12.009.
Hellsten, I., & Leydesdorff, L. (in press). Automated Analysis of Topic-Actor Networks on Twitter: New approach to the analysis of socio-semantic networks. *Journals of the Association for Information Science and Technology*. doi: 10.1002/asi.24207.
Holmberg, K., Bowman, T. D., Haustein, S., & Peters, I. (2014). Astrophysicists' Conversational Connections on Twitter. *PLOS ONE, 9*(8), e106086. doi: 10.1371/journal.pone.0106086.
Kamada, T., & Kawai, S. (1989). An algorithm for drawing general undirected graphs. *Information Processing Letters, 31*(1), 7-15. doi: 10.1016/0020-0190(89)90102-6.
Leydesdorff, L., & Zhou, P. (2007). Nanotechnology as a field of science: Its delineation in terms of journals and patents. *Scientometrics, 70*(3), 693-713. doi: 10.1007/s11192-007-0308-0.
Mas-Bleda, A., & Thelwall, M. (2016). Can alternative indicators overcome language biases in citation counts? A comparison of Spanish and UK research. *Scientometrics, 109*(3), 2007-2030. doi: 10.1007/s11192-016-2118-8.
McGill, G. S. a. M. J. (1983). *Introduction to Modern Information Retrieval*. Auckland: McGraw-Hill.
Moed, H. F. (2017). *Applied Evaluative Informetrics*. Heidelberg, Germany: Springer.
Mohammadi, E., Thelwall, M., & Kousha, K. (2016). Can Mendeley bookmarks reflect readership? A survey of user motivations. *Journal of the Association for Information Science and Technology, 67*(5), 1198-1209. doi: 10.1002/asi.23477.
National Information Standards Organization. (2016). *Outputs of the NISO Alternative Assessment Metrics Project*. Baltimore, MD, USA: National Information Standards Organization (NISO).
Pearce, W., Holmberg, K., Hellsten, I., & Nerlich, B. (2014). Climate Change on Twitter: Topics, Communities and Conversations about the 2013 IPCC Working Group 1 Report. *Plos One, 9*(4). doi: 10.1371/journal.pone.0094785.
Pooladian, A., & Borrego, Á. (2016). A longitudinal study of the bookmarking of library and information science literature in Mendeley. *Journal of Informetrics, 10*(4), 1135-1142. doi: 10.1016/j.joi.2016.10.003.
Priem, J., Taraborelli, D., Groth, P., & Neylon, C. (2010). Altmetrics: a manifesto. Retrieved March 28, from http://altmetrics.org/manifesto/
Robinson-García, N., van Leeuwen, T. N., & Ràfols, I. (2017). Using almetrics for contextualised mapping of societal impact: from hits to networks. Retrieved April 1, 2017, from https://papers.ssrn.com/sol3/papers.cfm?abstract_id=2932944
Sankar, S. A. (2015). Tweets Do Measure Non-Citational Intellectual Impact. *International Trends in Library and Information Technology (ITLIT), 2*(2).
Shema, H., Bar-Ilan, J., & Thelwall, M. (2014). Do blog citations correlate with a higher number of future citations? Research blogs as a potential source for alternative metrics. *Journal of the Association for Information Science and Technology, 65*(5), 1018-1027. doi: 10.1002/asi.23037.
Sugimoto, C. R., Work, S., Larivière, V., & Haustein, S. (2016). Scholarly use of social media and altmetrics: a review of the literature. Retrieved April, 5, 2017, from https://arxiv.org/abs/1608.08112

Sugimoto, C. R., Work, S., Larivière, V., & Haustein, S. (2017). Scholarly use of social media and altmetrics: A review of the literature. *Journal of the Association for Information Science and Technology, 68*(9), 2037-2062. doi: 10.1002/asi.23833.

Taylor, M. (2013). Towards a common model of citation: some thoughts on merging altmetrics and bibliometrics. *Research Trends*(35), 19-22.

Thelwall, M., & Kousha, K. (2015). Web Indicators for Research Evaluation. Part 2: Social Media Metrics. *Profesional De La Informacion, 24*(5), 607-620. doi: 10.3145/epi.2015.sep.09.

Thelwall, M., Kousha, K., Dinsmore, A., & Dolby, K. (2016). Alternative metric indicators for funding scheme evaluations. *Aslib Journal of Information Management, 68*(1), 2-18. doi: doi:10.1108/AJIM-09-2015-0146.

Vainio, J., & Holmberg, K. (2017). Highly tweeted science articles: who tweets them? An analysis of Twitter user profile descriptions. *Scientometrics, 112*(1), 345–366. doi: 10.1007/s11192-017-2368-0.

van Honk, J., & Costas, R. (2016). *Integrating context in Twitter metrics: Preliminary investigation on the possibilities of hashtags as an altmetric resource*. Paper presented at the Altmetrics 16, Bucharest. www.altmetrics.org/wp-content/uploads/2016/09/altmetrics16_paper_5.pdf

Wacholder, N. (2011). Interactive Query Formulation. *Annual Review of Information Science and Technology, 45*, 157-196. doi: 10.1002/aris.2011.1440450111.

Wouters, P., Thelwall, M., Kousha, K., Waltman, L., de Rijcke, S., Rushforth, A., & Franssen, T. (2015). *The Metric Tide: Literature Review (Supplementary Report I to the Independent Review of the Role of Metrics in Research Assessment and Management)*. London, UK: Higher Education Funding Council for England (HEFCE).

Wouters, P., Zahedi, Z., & Costas, R. (2018). Social media metrics for new research evaluation. *arXiv e-prints*. Retrieved from https://ui.adsabs.harvard.edu/\#abs/2018arXiv180610541W

Xia, F., Su, X., Wang, W., Zhang, C., Ning, Z., & Lee, I. (2016). Bibliographic Analysis of Nature Based on Twitter and Facebook Altmetrics Data. *PLOS ONE, 11*(12), e0165997. doi: 10.1371/journal.pone.0165997.